# Melting curve of black phosphorus: evidence for a solid-liquid-liquid triple point

*Hermann Muhammad[1], Mohamed Mezouar[1,\*], Gaston Garbarino[1], Laura Henry[2], Tomasz Poręba[1], Matteo Ceppatelli[3,4], Manuel Serrano-Ruiz[4], Maurizio Peruzzini[4], Frédéric Datchi[5]*

[1]European Synchrotron Radiation Facility (ESRF), 71, Avenue des Martyrs, Grenoble, France

[2]Synchrotron SOLEIL, 91192, Gif-sur-Yvette, France

[3]LENS, European Laboratory for Non-linear Spectroscopy, Via N. Carrara 1, I-50019 Sesto Fiorentino, Firenze, Italy

[4]ICCOM-CNR, Institute of Chemistry of OrganoMetallic Compounds, National Research Council of Italy Via Madonna del Piano 10, I-50019 Sesto Fiorentino, Firenze, Italy

[5]Institut de Minéralogie, de Physique des Milieux Condensés et de Cosmochimie (IMPMC), Sorbonne Université, CNRS UMR 7590, MNHN, 4 place Jussieu, F-75005 Paris, France

**Abstract**

Black phosphorus (bP) is a crystalline material that can be seen as ordered stackings of two-dimensional layers, which lead to outstanding anisotropic physical properties. The knowledge of its pressure-temperature (*P-T*) phase diagram, and in particular, the slope and location of its melting curve is fundamental for better understanding the synthesis and stability conditions of this important material. Despite several experimental studies, important uncertainties remain in the determination of this melting curve. Here we report accurate melting points measurements, using *in situ* high-temperature and high-pressure high-resolution synchrotron x-ray diffraction. In particular, we have employed an original and accurate pressure and temperature metrology based on the unique anisotropic *P-T* response of bP, that we used as sensor for the simultaneous determination of pressure and temperature up to 5 GPa and 1700 K. We confirmed the existence of and located a solid-liquid-liquid triple point at the intersection of the low- and high-pressure melting curves. Finally, we have characterized the irreversibility of the transformation in the low-pressure regime below 1 GPa, as the low-density liquid does not crystallize back to bP but into red phosphorus on temperature quenching.

**Introduction**

Phosphorus exhibits a very rich polymorphism under ambient [1] and non-ambient pressure-temperature (*P,T*) conditions [2]. The structure and physico-chemical properties of the numerous allotropes of phosphorus have been extensively studied using a variety of experimental and computational methods [2-14]. Three main allotropes are stable or metastable under ambient *P,T* conditions: white (wP), red (rP) and black phosphorus (bP). BP, the most stable allotrope [1], has been firstly prepared under high-pressure, high-temperature conditions using rP as starting material [15]. It features a layered structure similar to graphite, giving rise to strong covalent intralayer bonds and weak van der Waals interlayer interactions [14, 16-17]. The *P,T* phase diagram of bP exhibits unique features. At ambient *P* and high *T*, due to its exceptional anisotropic crystalline structure [14, 18-22], bP begins to thermally decompose at 600 K [23-24] and, in contrast with most elemental solids, does not melt or sublimate. The lowest *P-T* conditions at which melting of bP has been reported are 0.3 GPa and 1123 K. Another fundamental peculiarity that bP only shares with orthorhombic sulfur [25] is that it exhibits a first-order phase transition in its liquid state [26] between a low-density liquid composed of P4 molecular units [26-27] and a high-density liquid with a local atomic arrangement resembling that of bP [27]. The melting curve of bP has been investigated by differential thermal analysis (DTA) [28] and *in situ* energy dispersive x-ray diffraction [29-31] in piston-cylinder apparatus [32] and multi-anvil presses [33]. However, contradicting results have been reported regarding its slope and position.

Here, based on an original *P,T* metrology, we accurately determined the melting curve position and slope using *in situ* high-resolution synchrotron x-ray diffraction performed under high-pressure and -temperature conditions. Besides providing an accurate melting curve of bP between 0 and 5.5 GPa and between 300 and 1700 K, we have established the position of the solid-LDL-HDL triple point (SLHTP) and shown the irreversibility of the transformation of low-density liquid P into rP in the low-pressure regime below 1 GPa.

**Method**

All experimental runs were carried out at the ID27 high-pressure x-ray diffraction beamline of the European Synchrotron Radiation Facility (Grenoble, France) [34]. A VX5 Paris-Edinburgh press [35-37] has been employed as pressure device. The sample consisted of a high purity (99.999+%) powder of black phosphorus (bP) produced following the synthesis method described in [38]. It was confined in a sample assembly optimized for accurate melting point determination [37, 39]. A diamond cylinder of 1 mm inner diameter, 1.5 mm outer diameter and 1 mm height was used as inert sample container. The excellent thermal conductivity of diamond ensured very low temperature gradients within the x-ray probed volume. The diamond capsule was sealed using two chemically inert hexagonal boron nitride caps that served as soft pressure medium. The high temperatures up to 1800 K were generated using a high resistivity cylindrical graphite heater supplied with direct current from a delta power supply. This heater was inserted in an x-ray transparent boron-epoxy gasket that served as thermal and electrical insulator [39]. *In situ* monochromatic x-ray diffraction experiments have been carried out in transmission geometry either at 20.0 keV ($\lambda$=0.6199 Å) or at 33.169 keV ($\lambda$=0.3738 Å) to cover a large 2 theta scattering angle from 3 to 25 degrees. Two-dimensional diffraction patterns were collected using a MAR165 CCD detector from MAR Research or an EIGER2 9M pixel detector from DECTRIS. Typical exposure time of 10 seconds was sufficient to collect high-quality XRD patterns. A high efficiency multichannel collimator (MCC) [40] was used to remove most of the parasitic elastic and inelastic x-ray signal coming from the sample environment (graphite heater and boron-epoxy gasket). The sample to detector distance, detector tilt angles and beam centre were accurately determined using $LaB_6$ powder as standard. The two-dimensional XRD images were integrated using the PyFAI software [41] as implemented in the DIOPTAS suite [42]. The resulting one-dimensional diffraction patterns were analysed using the GSAS software [43] to refine the unit-cell parameters and volume of bP by Le Bail [44] extraction of *d*-spacings using a pseudo-Voigt peak shape function. The main objective of this study is to accurately establish the position and slope of the melting curve of bP. Therefore, it is of crucial importance to accurately determine the pressure $P$ and temperature $T$ at which the XRD measurements have been performed. Here, we have established an original and accurate $P,T$ metrology that is detailed in ref. [22]. In short, it is based on exploiting the highly anisotropic character of bP originating from its particular layered structure. Indeed, in its $P,T$ stability field, bP is quasi-incompressible along the $a$-axis, exhibiting a high directional thermal expansion coefficient in this direction ($\alpha_a$= 6.46(6)·10$^{-6}$ K$^{-1}$). This remarkable feature makes it the only element that can be employed as simultaneous $P,T$ sensor. In practice, while the $a$ lattice parameter of bP can be determined from the (k00) Bragg reflections, the temperature can then be derived using its linear T dependence established in [22]:

$$T(K) = \frac{a - a_0}{\alpha_a} + 300$$

where $a$ and $a_0$ are the values of the $a$ lattice parameter at a given pressure and temperature, and ambient conditions, respectively, and $\alpha_a$ is the corresponding directional thermal expansion coefficient. The pressure is then derived from the third-order Birch-Murnaghan equation of state [45] of bP. This method enables the determination of $P$ and $T$ with an accuracy of ± 25 K and 0.1 GPa in the entire $P,T$ stability field of bP up to 5 GPa and 1700 K.

Another fundamental aspect for this type of studies resides in the use of an unambiguous melting criterion. Here we have employed a standard and well-established criterion [46-48]: the x-ray signature associated with the loss of crystalline order that occurs at melting. As shown in Fig. 1, through the use of a multi-channel collimator (MCC) [49], high quality diffraction patterns were obtained for solid and liquid phosphorus. The transition between these two states is evidenced by the disappearance of the Bragg reflections of bP that gives rise to a strong diffuse x-ray signal when the melting line is crossed. This happens during melting either towards the low-density liquid (LDL) when the pressure is reduced at constant temperature (Fig. 1a) or towards the high-density liquid (HDL) when the temperature is increased at constant pressure (Fig. 1b).

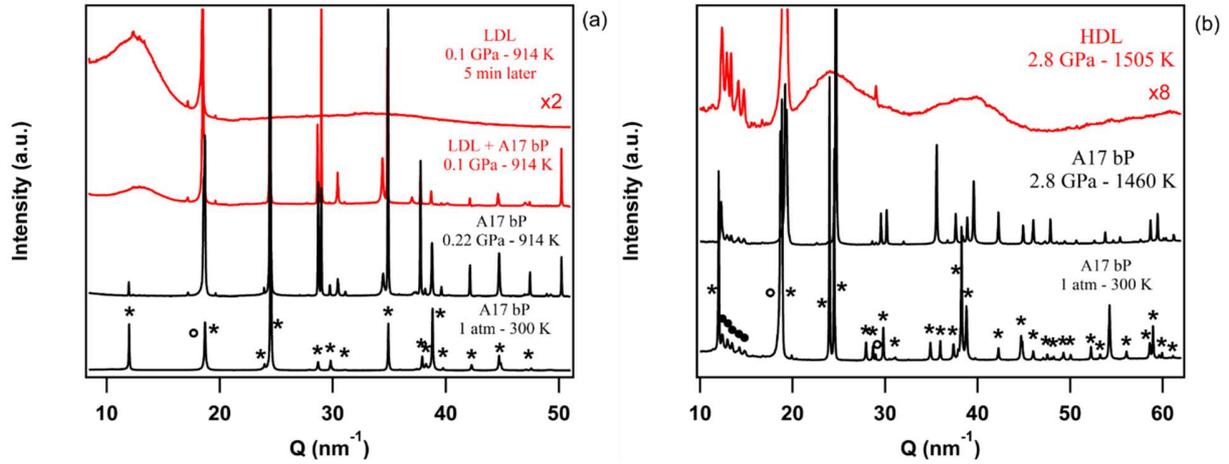

**Figure 1.** Melting criteria. XRD diagrams of bP collected on decompression at constant temperature (914 K) toward the low-density liquid (a), and increasing temperature at constant pressure (2.8 GPa) toward the high-density liquid (b). The solid-liquid coexistence is evidenced in the left panel (a). The black asterisks, black filled circles, and empty white circles indicate the Bragg reflections of bP, the boron-epoxy gasket, and the hexagonal boron nitride container, respectively.

**Results and discussion**

Typical $P,T$ pathways in the phase diagram of bP in the range 0.1-5.3 GPa and 300-1800 K are shown in Fig. 2. A new fresh loaded sample was employed for each of them to ensure the high quality of the collected datasets and control their reproducibility. XRD patterns were systematically collected along 19 distinct $P,T$ pathways leading to the determination of 19 melting points. It is worth nothing that we were able to collect data points at very low pressure down to 0.10 ($\pm$ 0.03) GPa, where solid-liquid coexistence has been observed. Low-pressure melting points determination is challenging due to the mechanical instability of bP near ambient pressure [23-24]. In practice, to reach the melting curve, the temperature was increased by small increments of approximately 30 K at constant pressure or the pressure was finely reduced at constant temperature by steps of 0.1 GPa to bracket the melting temperatures and pressures with an accuracy of $\pm$ 25 K and 0.1 GPa. At pressures below 1.2 GPa, the melting of bP towards the low-density liquid has been clearly assessed by the appearance of a strong diffuse x-ray signal which exhibits a first sharp diffraction peak (FSDP) at a Q value of 13 nm$^{-1}$ (Fig.1a). As reported elsewhere [26-27], this FSDP is associated to the medium-range correlations between the P$_4$ tetrahedra that are present in the LDL. At higher pressures (P>1.2 GPa), the melting occurs toward the high-density liquid (HDL) which exhibits two distinct maxima in the interference function S(Q) at 2.3 and 40 nm$^{-1}$ (Fig. 1b) that are unambiguous signatures of the HDL [27]. This enabled the accurate location of the solid-LDL-HDL triple point (SLHTP) at 1.35 ± 0.15 GPa and 1350 ± 25 K. It is worth noting that the remaining Bragg reflections present at low Q (Q < 30 nm$^{-1}$) are assigned to the sample container material (boron-epoxy gasket, graphite heater and h-BN capsule) that are not completely filtered out by the MCC [49].

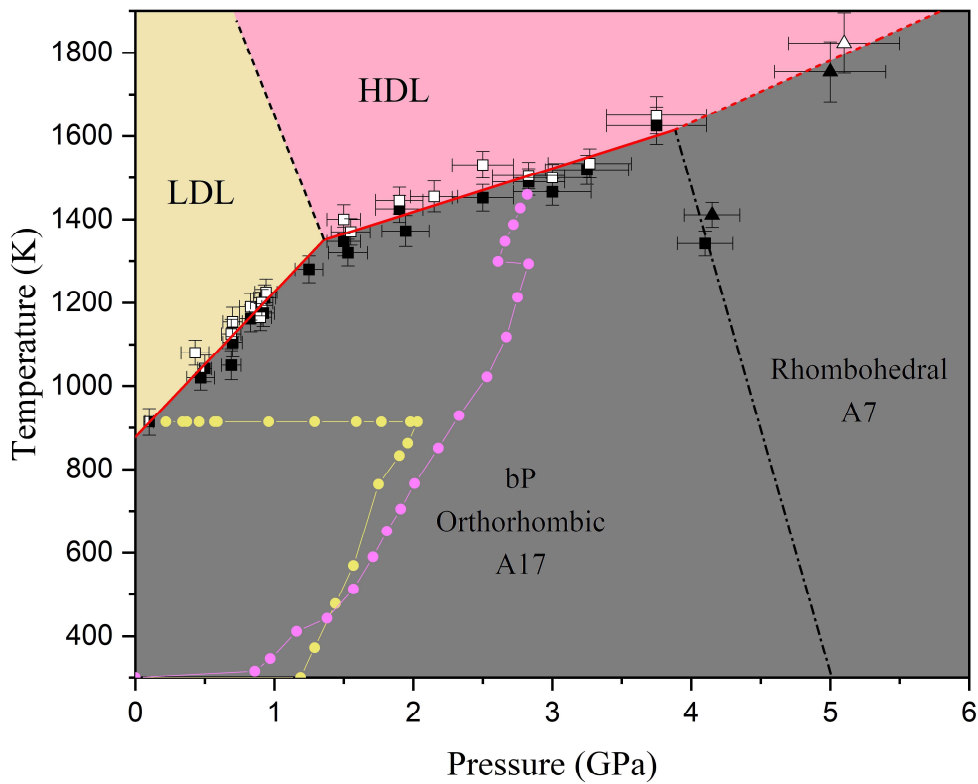

**Figure 2.** Melting curve of bP. Plain and empty squares denote respectively the solid and liquid state. Black squares and triangles indicate the A17 and A7 phase of solid bP, respectively. Two typical P-T pathways are highlighted (yellow and pink filled circles and lines).

In agreement with previous works [28-31], we did not observe any structural phase transition prior to the melting of the A17 phase of phosphorus. As shown in Fig. 1a, we instead evidenced the coexistence between the A17 phase and the LDL at 0.1 GPa and 914 K. This confirms that bP is the thermodynamically stable allotrope in the 0 to 5 GPa range. As reported previously, at higher pressure bP becomes energetically less favorable and transforms to the A7 rhombohedral phase (*R-3m* space group). In the course of this experiment, we could determine one melting point in the stability field of the A7 polymorph at 5.05 GPa and 1780 K. From this information and the location of 2 additional points on the A17 to A7 solid-solid transition line at 4.12 GPa and 1375 K (this work) and 5 GPa and 300 K [2], we could infer the position of the A17-A7-HDL triple point at 3.88± 0.15 GPa and 1615± 25 K.

| Pressure (GPa) | Temperature (K) | Initial and Final Phase |
|---|---|---|
| 0.10 (3) | 914 (25) | A17 to LDL |
| 0.45 (10) | 1050 (25) | A17 to LDL |
| 0.50 (5) | 1042 (25) | A17 to LDL |
| 0.69 (7) | 1088 (30) | A17 to LDL |
| 0.70 (7) | 1129 (30) | A17 to LDL |
| 0.83 (8) | 1176 (30) | A17 to LDL |
| 0.92 (8) | 1205 (30) | A17 to LDL |
| 0.93 (8) | 1200 (30) | A17 to LDL |
| 1.25 (10) | 1330 (30) | A17 to LDL |
| 1.50 (12) | 1374 (30) | A17 to HDL |
| 1.54 (12) | 1355 (30) | A17 to HDL |
| 1.90 (15) | 1435 (30) | A17 to HDL |
| 2.05 (15) | 1414 (35) | A17 to HDL |
| 2.50 (20) | 1492 (30) | A17 to HDL |
| 2.83 (25) | 1498 (30) | A17 to HDL |
| 3.00 (28) | 1483 (30) | A17 to HDL |
| 3.26 (30) | 1527 (35) | A17 to HDL |
| 3.75 (35) | 1637 (45) | A17 to HDL |
| 5.05 (40) | 1788 (70) | A7 to HDL |

**Table 1.** *P,T* location of the melting points of bP and the initial and final solid and liquid phase.

The melting curve of bP and the *P,T* location of the individual melting points are presented in Figure 2 and Table 1, respectively. The lowest pressure melting point has been observed near the ambient pressure axis at 0.1 GPa and 914 K. From the extrapolation of the melting curve, the ambient pressure melting point is estimated at $880 \pm 15$ K in good agreement with the extrapolation to zero reported in [28,30-31]. We observed a linear pressure increase in melting temperature with two distinct pressure regimes. Indeed, the slope of the melting curve is suddenly and significantly modified at the SLHTP. Its value is multiplied by almost a factor of 3 varying from $105 \pm 12$ K·GPa$^{-1}$ to $348 \pm 21$ K·GPa$^{-1}$. This abrupt change is likely due to the large local atomic order difference between the LDL and HDL mentioned above. From the Clausius-Clapeyron relation [49,50], significant differences in the associated melting entropy and volume between the low- and high-pressure regimes (below and above the SLHTP, respectively) are expected. As already mentioned, the SLHTP was located at the intersection of the low- and high-pressure melting lines at $1.35 \pm 0.15$ GPa and $1350 \pm 25$ K.

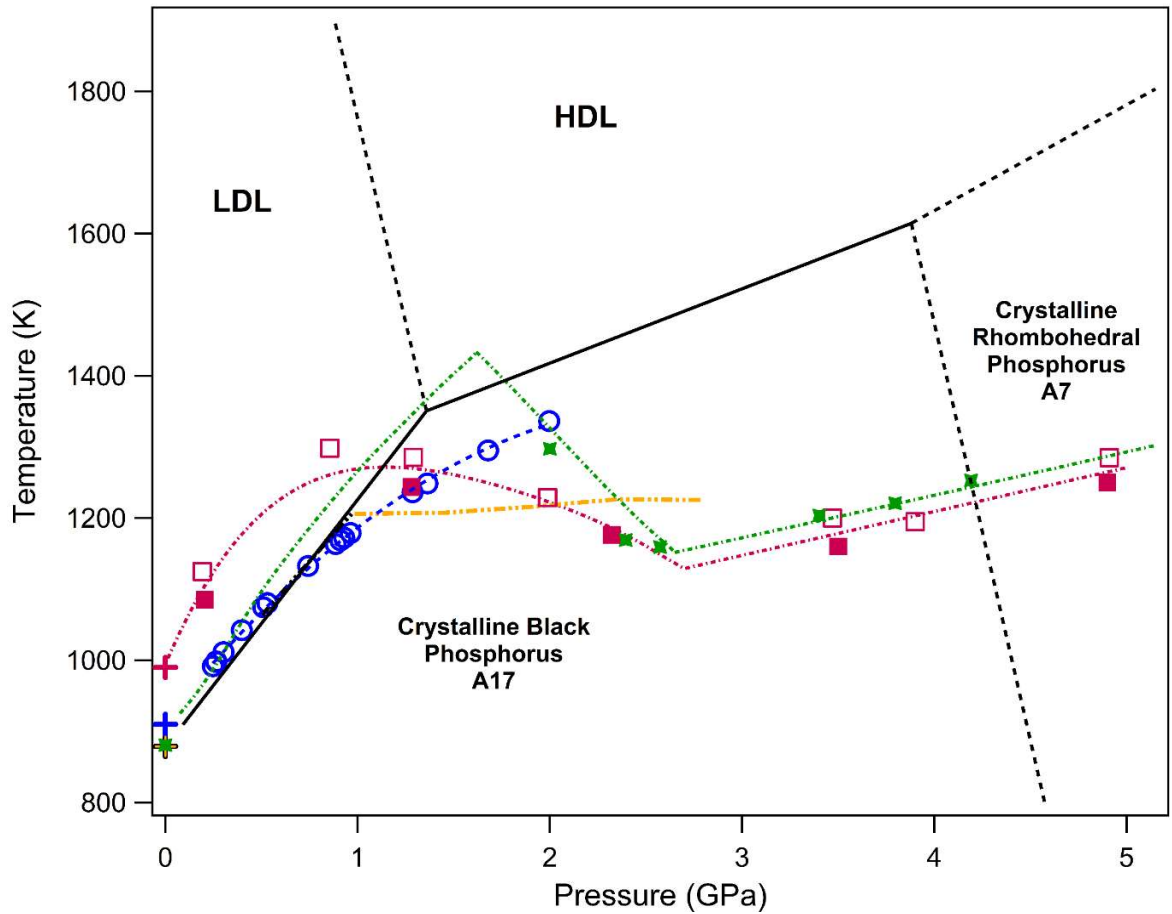

**Figure 3.** Comparison between the melting curve of bP from this work (black line) and the melting curve from Akahama et al [29] (red dashed line and symbols), Marani and Guarise [28] (blue dashed line and symbols), Mizutani et al [30] (yellow dashed line) and Solozhenko et al [31] (green dashed line and symbols). Plain and empty squares denote respectively the solid and liquid state. The colored crosses indicate the melting temperatures obtained by extrapolation of the melting curves to ambient pressure.

A comparison of the melting curve of bP determined in this study with literature data is shown in Figure 3. Very significant appear between these different works and the one presented here. These are most probably due to a variety of experimental problems which are discussed in the following. The melting curve reported by Marani and Guarise [28] was determined by differential thermal analysis (DTA). It agrees fairly well with our results up to ~1 GPa but deviates from them at higher pressures. Indeed, no break in slope is highlighted and the melting curve of Marani and Guarise appears to follow a monotonous quadratic variation. Moreover, no data was collected above 2.3 GPa, which does not allow for a complete comparison. The deviation observed above 1 GPa is probably due to the inaccuracy of the temperature measurement of the DTA method at high pressure. Indeed, this technique does not consider the pressure effect on the temperature probe (thermocouple) which are nevertheless significant [51]. The three other reported works [29-31] were carried out using the same *in situ* energy dispersive x-ray diffraction in multi-anvil presses methodology [33]. However, these studies have led to widely divergent results. Akahama et al. [29] reported a maximum of the melting curve around 1 GPa, followed by a reduction of the melting temperature toward the triple point between the HDL, orthorhombic bP and the rhombohedral A7 phase located around 2.7 GPa and 1100 K. In a more recent work, Solozhenko et al. [31], have obtained qualitatively similar results but with a much sharper slope of the melting curve in the low-and high-pressure regimes despite the absence of data points near the SLHTP. In Mizutani et al. [30], in agreement with the current work, a break in the slope of melting is observed curve around 1 GPa. However, the reported slope in the high-pressure regime strongly differs from our result by one order of magnitude ($dT/dP = 105 \pm 12$ K·GPa$^{-1}$ and $dT/dP = 15$ K·GPa$^{-1}$, in this work and Mizutani et al. [30], respectively). The diverging results could be explained by 3 main reasons. As for the DTA work of Marani and Guarise [28] a thermocouple was employed as temperature sensor and the significant pressure effect was neglected. The second potential source of imprecision originates from the employed melting criterion based on the energy dispersive x-ray diffraction (EDX) method. Indeed, in the EDX method, the x-ray detection is carried out using a single-point detector [52] that intercepts only a very small fraction of the photons diffracted by the sample. At high temperature and more particularly near the melting point, it is well known [53-54] that materials undergo strong recrystallization effects. This, in turn, transform a randomly oriented x-ray powder patterns into highly oriented diffraction spots that occupy a much smaller fraction of the Debye-Scherrer rings. Contrary to the present work, the employed melting criterion in [31] is solely based on the disappearance of the Bragg reflections. This could lead to an incorrect determination of melting as it can be confused with recrystallization. The third problem resides in the low number of collected data points in the previous studies. As a comparison, in the references [29] and [31] less than 5 melting points were determined compared to 19 in the present study. Another fundamental advantage here is related to the employed metrology. As explained in the method section and detailed in [22], we used the anisotropic properties of the sample itself to simultaneously extract $P$ and $T$ with very good precision. As the bP sample is used as $P,T$ sensor, this method also guarantees the absence of pressure and temperature gradients, which is not the case with the other methods mentioned above. Indeed, in multi-anvil presses [33], the pressure calibrant and the temperature probe is located at a finite distance (typically 1 mm) from the sample, which inevitably results in larger systematic errors.

In complement to determining the melting curve slope and position, we have studied the reversibility of the transition by *in situ* characterization of the crystal structure of the solids recovered after temperature quenching of the melt. As shown in Fig. 4a and 4b, at pressures above the SLHTP, a behavior in agreement with a previous report [55] is observed. After solidification on decreasing the temperature, all the apparent Bragg reflections have been assigned to bP, thus confirming the reversibility of the transformation in this pressure regime. However, at pressures below the SLHTP, as briefly reported in [29], the phase transition is irreversible. Indeed, as determined from Le Bail refinement of the XRD pattern, the recovered solid adopts a structure compatible with that of red phosphorus (rP). Fibrous rP is a known triclinic polymorph of phosphorus with space group a*P*42 [56]. The refined unit-cell parameters are listed in Table 2 with the literature data. The notable mismatch with the literature suggests that although the refinement is compatible with a*P42*, the only known crystallographic structure for crystalline rP, this refinement may not be conclusive. The reversibility of melting above the SLHTP suggests a proximity of the local order in HDL and bP, while its irreversibility below the SLHTP shows that the local atomic arrangement in LDL differs strongly from that in bP. Indeed, in such case we may expect a large surface energy between LDL and bP, which makes the nucleation of bP unfavored compared to rP. A similar case has been reported in water. When it is supercompressed at 1.7 GPa, it crystallizes into the metastable ice VII rather than the stable ice VI due to a smaller liquid/ice VII surface energy [57]. A more detailed study of the quenched solid and of its melting line should however be undertaken to better understand the irreversible character of the melting of bP below the triple point, and will be the subject of future works. The different nature of the quenched products of LDL and HDL indirectly confirms that the local atomic arrangement in these two liquid phases is substantially different.

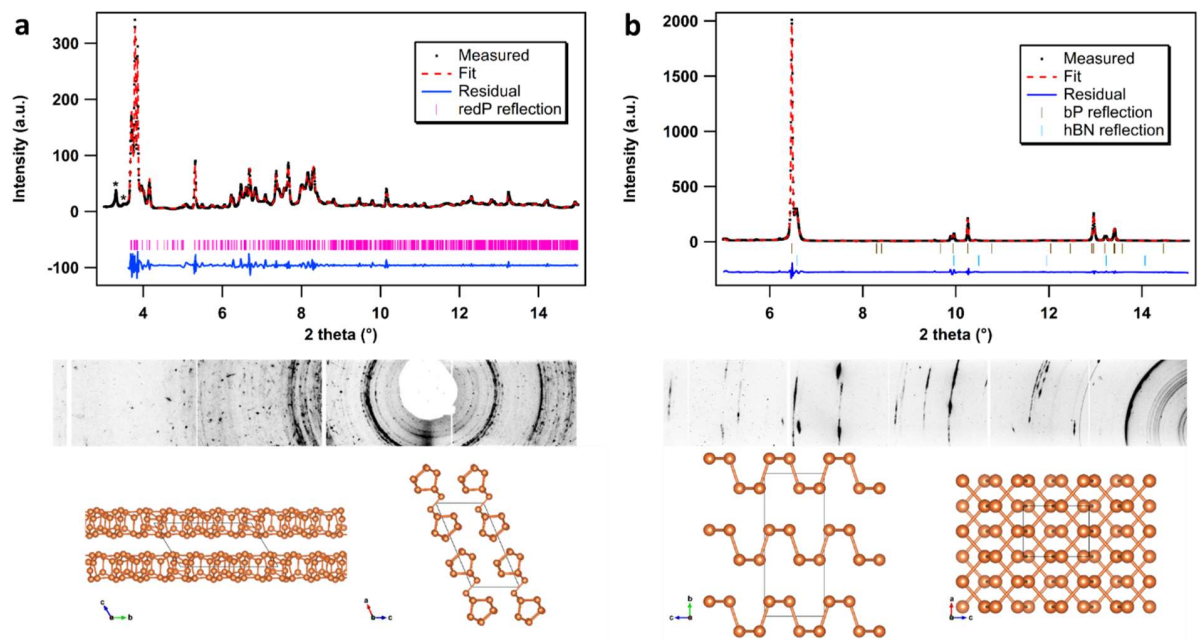

**Figure 4.** Upper panels: Diffraction patterns of the quenched products from the LDL (a) and from the HDL (b) and corresponding 3D crystalline structures in different crystallographic directions. Lower panels: (a) XRD image plates and Le Bail refinement of rP collected at ambient conditions after temperature and pressure quenching. (b) XRD image and Le Bail refinement of bP collected at ambient conditions after temperature and pressure quenching.

| Exp | a (Å) | b (Å) | c (Å) | α (°) | β (°) | γ (°) | V (Å³) | Conditions |
|---|---|---|---|---|---|---|---|---|
| This work | 12.291(10) | 13.043(6) | 7.168(7) | 117.42(4) | 108.60(7) | 95.04(6) | 928.4(13) | Ambient |
| Ruck et al. [56] | 12.198(8) | 12.986(8) | 7.075(7) | 116.99(7) | 106.31(7) | 97.91(7) | 911.25 | Ambient |

**Table 2.** Refined unit cell parameters of aP42 red phosphorus from this study and from the literature.

**Conclusion**

In this study, we employed *in situ* high-resolution x-ray diffraction to accurately determine the melting curve of bP in an extended *P,T* range up to 5 GPa and 1700 K, and to characterize the reversibility of the liquid to solid transformation. The melting curve was determined with unparalleled accuracy using an original metrology which takes full advantage of the exceptional anisotropic properties of bP [2,22]. This enabled the simultaneous and accurate measurement of *P* and *T* using the sample itself as *P,T* sensor, insuring the absence of significant *P,T* gradients in the sample volume probed by the x-ray beam. Based on this experimental approach, two distinct regimes of melting, corresponding to the low- and high- density liquid were observed, allowing the precise determination of the triple point between the LDL, HDL and crystalline bP in the *P,T* phase diagram of P. The very contrasting slope of the melting curve indirectly confirms the striking difference in the local atomic arrangement of the LDL and HDL, as evidenced by the irreversible nature of the transformation in the low-pressure regime, where the low-density liquid obtained by melting bP irreversibly crystallizes into rP. This suggests that the local order in the LDL, currently considered as composed of $P_4$ molecular units, could have instead a much more complex structure, stimulating further systematic structural studies of the LDL and of the transition line between the LDL and HDL.

**AUTHOR INFORMATION**

**Corresponding Author**

**Mohamed Mezouar** - European Synchrotron Radiation Facility (ESRF), 71 Avenue des Martyrs, Grenoble, France; Email: mezouar@esrf.fr; Phone: +33 (0)4 76 88 25 15
**Author Contributions**

The original idea was defined by MM. Experiments were performed by HM, GG, TP, MC, MSD and MM with equal contributions. The data were analyzed and the figures were produced by HM with contributions from all the co-authors. The manuscript was written by MM with contributions from all authors. All authors have given approval to the final version of the manuscript.

**Funding Sources**

Thanks are expressed to EC through the European Research Council (ERC) for funding the project PHOSFUN ''Phosphorene functionalization: a new platform for advanced multifunctional materials'' (Grant Agreement No. 670173) through an ERC Advanced Grant, and the Agence Nationale de la Recherche (ANR) for financial support under grant ANR-21-CE30-0032-01 (LILI).

**Notes**

There are no conflicts to declare.

**ACKNOWLEDGMENT**

We acknowledge the ESRF for the provision of beamtime on the High-Pressure beamline ID27. Thanks are expressed also to the projects "GreenPhos – alta pressione" (CNR), HP-PHOTOCHEM (Fondazione Cassa di Risparmio di Firenze) and PRIN 2017 KFY7XF FERMAT "FastElectRon dynamics in novel hybrid-2D MATerials" (MUR).